\documentclass[a4paper,preprint,showpacs,amssymb,pre,superscriptaddress]
              {revtex4}
\usepackage{graphicx}

\begin{document}

\title{Stochastic resonance between dissipative structures in a
bistable noise-sustained dynamics}

\author{B. von Haeften}
\author{G. Iz\'us\cite{conicet}}
\author{S. Mangioni\cite{conicet}}
\author{A. D. S\'anchez\cite{conicet}}
\affiliation{Departamento de F\'{\i}sica, FCEyN,
             Universidad Nacional de Mar del Plata\\
         De\'an Funes 3350, 7600 Mar del Plata, Argentine}
\author{H. S. Wio\cite{conicet}}
\email{wio@cab.cnea.gov.ar} \affiliation{Grupo de F\'{\i}sica
Estad\'{\i}stica,
             Centro At\'omico Bariloche and Instituto Balseiro, \\
             8400 San Carlos de Bariloche, Argentine}
\affiliation{Departament de F\'{\i}sica,
         Universitat de les Illes Balears and IMEDEA \\
         E-07122 Palma de Mallorca, Spain}

\begin{abstract}
We study an extended system that without noise shows a monostable
dynamics, but when submitted to an adequate multiplicative noise,
an effective bistable dynamics arise. The stochastic resonance
between the attractors of the \textit{noise-sustained dynamics} is
investigated theoretically in terms of a two-state approximation.
The knowledge of the exact nonequilibrium potential allows us to
obtain the output signal-to-noise ratio. Its maximum is predicted
in the symmetric case for which both attractors have the same
nonequilibrium potential value.
\end{abstract}


\maketitle

\section{Introduction}

During the last few decades a wealth of research results on
fluctuations or noise have lead us to the recognition that in many
situations noise can actually play a constructive role that
induces new ordering phenomena. Some examples are stochastic
resonance in zero-dimensional and extended systems
\cite{RMP,extend1,extend2,extend2b,extend3a,extend3b},
noise-induced transitions \cite{lefev}, noise-induced phase
transitions \cite{nipt1,nipt2}, noise-induced transport
\cite{Ratch2,Ratch3,nipt3,nipt4}, noise-sustained patterns
\cite{ga93,nsp,ber}, noise-induced limit cycle \cite{mangwio},
etc.

The phenomenon of \textit{stochastic resonance} (SR)--- namely,
the \textit{enhancement} of the output signal-to-noise ratio (SNR)
caused by injection of an optimal amount of noise into a nonlinear
system--- stands as a puzzling and promising cooperative effect
arising from the interplay between \textit{deterministic} and
\textit{ random} dynamics in a \textit{nonlinear} system.  The
broad range of phenomena--- drawn from almost every field in
scientific endeavor--- for which this mechanism can offer an
explanation has been put in evidence by many reviews and
conference proceedings. See Ref.\cite{RMP} and references there to
scan the state of the art.

Most of the phenomena that could possibly be explained by SR occur
in \textit{extended} systems: for example, diverse experiments
were carried out to explore the role of SR in sensory and other
biological functions \cite{biol} or in chemical systems
\cite{sch}. These were, together with the possible technological
applications, the motivation to many recent studies showing the
possibility of achieving an enhancement of the system response by
means of the coupling of several units in what conforms an
\textit{extended medium}
\cite{extend1,otros,extend2,extend3a,extend3b}, or analyzing the
possibility of making the system response less dependent on a fine
tuning of the noise intensity, as well as different ways to
control the phenomenon \cite{claudio,nos3}.

In some previous papers \cite{extend2,extend2b,extend3a,extend3b}
we have studied the stochastic resonant phenomenon in extended
systems for the transition between two different patterns, and
exploiting the concept of \textit{nonequilibrium potential}
\cite{GR,I0}. The nonequilibrium potential is a special Lyapunov
functional of the associated deterministic system which for
nonequilibrium systems plays a role similar to that played by a
thermodynamic potential in equilibrium thermodynamics \cite{GR}.
Such a nonequilibrium potential, closely related to the solution
of the time independent Fokker-Planck equation of the system,
characterizes the global properties of the dynamics: that is
attractors, relative (or nonlinear) stability of these attractors,
height of the barriers separating attraction basins, and in
addition it allows us to evaluate the transition rates among the
different attractors.

In this work we analyze a new aspect of such a problem studying SR
between the attractors of the \textit{noise-sustained dynamics}
\cite{ga93,nsp}, that is: the same noise source that induces the
dynamics, induces the transitions among both structures, and
produces the stochastic resonant phenomenon. Some closely related
work correspond to the so called \textit{doubly stochastic
resonance} \cite{ZKSG1}, as well as to another previous work
\cite{MGOC} related with noise-induced phase transitions
\cite{nipt1,nipt2}. In both cases the authors have mainly resorted
to a standard mean-field approach, or to an estimate of the
effective potential, while here we obtain the exact form of the
noise-induced patterns (stable and unstable ones) as well as the
complete form of the nonequilibrium potential. In this way we can
obtain the transition rates and clearly quantify the SR phenomenon
by means of the SNR.

The organization of the paper is as follows. In section
\ref{model} we present the model and formalism to be used. After
that, we discuss in section \ref{sr:sec} the stochastic resonance
phenomenon between the homogeneous structure and the inhomogeneous
pattern. Finally, we present in section \ref{conc} some
conclusions and future perspectives.

\section{\label{model}The Model}

We consider a one-dimensional system, limited to the region $-L/2
\le x \le L/2$, described by the following deterministic equation
\begin{equation}
\label{determinista}
\partial_t \phi(x,t)=\partial_x(D(\phi) \partial_x \phi)+F(\phi),
\end{equation}
assuming Dirichlet boundary conditions (that is $\phi(\pm
L/2)=0$). This equation can be written in a variational form as
\begin{equation}
\partial_t \phi(x,t)=-\frac{1}{D(\phi)}\frac{\delta V[\phi]}{\delta
\phi(x)},
\end{equation}
where the potential
\begin{equation}
\label{pot}
V[\phi]=\int_{-L/2}^{L/2} dx \bigg \{ -\int_0^{\phi} d
\phi' D(\phi') F(\phi')+\frac{1}{2} \big[D(\phi)\partial_x \phi
\big]^2 \bigg \}
\end{equation}
is a Lyapunov functional (while $D(\phi)>0$)  for the
deterministic dynamics and it is essentially the logarithm of the
probability density of configuration when Eq.(\ref{determinista})
is perturbed by an additive source of spatiotemporal white noise.

The starting point of our stochastic analysis will be Eq.
(\ref{determinista}) with an additional multiplicative noise, in
the Stratonovich interpretation, given by
\begin{equation}
\label{partida}
\partial_t \phi(x,t)=-\frac{1}{D(\phi)}\frac{\delta V[\phi]}{\delta
\phi(x)}+ g(\phi ) \xi(x,t),
\end{equation}
where $\xi$ is a Gaussian noise with zero mean and correlation
$\langle \xi(x,t) \xi(x',t') \rangle=2 \epsilon
\delta(x-x')\delta(t-t')$, being $\epsilon$ the noise intensity.
For the coefficient of the noise term, $g(\phi )$, we adopt
\begin{equation}
\label{ruidomult} g(\phi ) = \frac{1}{\sqrt{D(\phi)}},
\end{equation}
in order to guarantee that the fluctuation-dissipation relation is
fulfilled \cite{kitara}.

As we are considering the Stratonovich interpretation, the
stationary solution of the associated Fokker-Planck equation can
be written as \cite{ibanyes}
\begin{equation}
\label{solucion} P_{st}[\phi] \sim \exp(-V_{eff}/\epsilon),
\end{equation}
where the effective potential $V_{eff}[\phi]$ is given by
\begin{equation}
V_{eff}[\phi]=V[\phi]-\lambda \int_{-L/2}^{L/2} dx \ln D(\phi).
\end{equation}
Here $\lambda$ is a renormalized parameter related to $\epsilon$
through $\lambda=\epsilon/(2 \Delta x)$ in a lattice
discretization, where $\Delta x$ is the lattice parameter
\cite{ibanyes}.

The extremes of $V_{eff}$ correspond to the stationary fixed
points of the noise-sustained dynamics. They can be computed from
the first variation of $V_{eff}(\phi)$ respect to $\phi$ equal to
zero, that is
\begin{equation}
\label{equivalente} \delta V_{eff} [\phi_{st}]=-\int_{-L/2}^{L/2}
D(\phi)[\partial_x (D(\phi) \partial_x \phi)+F_{eff}(\phi)] \delta
\phi(x)\, dx \bigg|_{\phi=\phi_{st}}=0,
\end{equation}
where
\begin{equation}
 \label{fundamental}
F_{eff}(\phi)=F(\phi)+\lambda
\frac{1}{D(\phi)^2}\frac{d}{d\phi} D(\phi)
\end{equation}
is the effective nonlinearity which drives the dynamics.

We consider the case of a monostable dynamics in absence of noise
\begin{equation}
\label{efe} F=-\phi^3+b\phi^2,
\end{equation}
and we adopt a model of field-dependent diffusivity which induces
an effective bistable dynamics. In particular we have chosen
\begin{equation}
\label{difu} D(\phi)=\frac{D_0}{1+h \phi^2 },
\end{equation}
($D_0,\, h >0$), that corresponds to have a larger diffusivity in
low density (low $\phi$) regions and a lower diffusivity in high
density (large $\phi$) ones. With this functional form,
$F_{eff}(\phi)$ in Eq. (\ref{fundamental}) results
\begin{equation}
F_{eff}=-\phi^3+b\phi^2-\frac{2\lambda h
\phi}{D_0}=\phi(\phi-\phi_1)(\phi_2-\phi),
\end{equation}
where $\phi_{1,2}$ depend on parameters, in particular on the
control parameter $\lambda$. It is worth noting here that in the
deterministic problem ($\lambda=0$) the reaction term is
monostable while, as we increase the noise intensity, the
effective nonlinear term $F_{eff}$ becomes bistable (within the
interval $0<\lambda<b^2 D_0/(8h)$) and finally, for $\lambda>b^2
D_0/(8h)$ becomes again  monostable (reentrance effect). Our
choice of $F$ and $D$ is one among a plenty of different forms for
the diffusivity leading to a transition from monostable to
bistable and inducing the SR phenomenon (see for instance the one
used in \cite{ibanyes}, that corresponds exactly to the inverse of
the present diffusion coefficient, i.e. $D(\phi)=D_0(1+h
\phi^2)$). Density-dependent diffusivities arise in a large
variety of systems modelled by reaction-diffusion equations
\cite{ddd0}. In biology, for instance, population dynamics is
usually driven by a diffusivity that depends on the local
population \cite{ddd1}. We can also find examples in physics, a
couple of them are in polymer physics (where the diffusion can
abruptly drop several orders of magnitude at the gelation point
\cite{ddd2}) and in diffusion of hydrogen in metals \cite{ddd3}.

A remarkable point is that $\phi=0$ is always a root of
$F_{eff}=0$ (see Fig. 1). This implies (from Eq.
(\ref{equivalente})) that $\phi(x)\equiv 0$ is an extremum of
$V_{eff}[\phi]$ for all values of $\lambda$. In what follows we
will call this structure $\phi_0$.

\begin{figure}
\centering
\includegraphics[width=10cm]{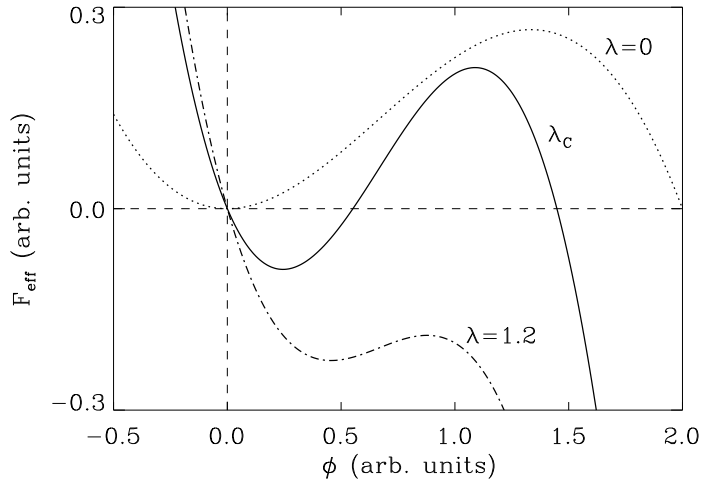}
\caption{Form of the nonlinearities for the deterministic case
($\lambda=0$), bistable case ($\lambda=\lambda_c \approx 0.8$) and
a monostable case ($\lambda=1.2$) in the reentrance region. The
vertical scale was changed in the deterministic case in order to
clarify the figure. The parameters used are: $D_0=1$, $h=1/2$ and
$b=2$. Note that $\phi=0$ remains as a root in all cases.}
\end{figure}

In order to obtain the non uniform extremes of the potential (and
also of the probability density) we must (numerically) solve
\begin{equation}
\frac{d}{dx}\bigg(D(\phi_{st}) \frac{d}{dx}
\phi_{st}\bigg)+F_{eff}(\phi_{st}) =0,
\end{equation}
for the stationary regimen profiles $\phi_{st}(x)$. This approach
allows us to found both, the stable and unstable solutions. To
analyze their stability we need to calculate $\delta ^2 V_{eff}$,
that defines a Sturm-Liouville problem, with orthogonality weight
$D(\phi_{st})$. From that analysis it results that $\phi_0$
(defined before) is stable for $\lambda>0$, and in the bistability
region we have two nonhomogeneous symmetric patterns: one unstable
$\phi_u$ (saddle) and one stable $\phi_s$. The typical form of
these patterns is illustrated in Fig. 2.

\begin{figure}
\centering
\includegraphics[width=10cm]{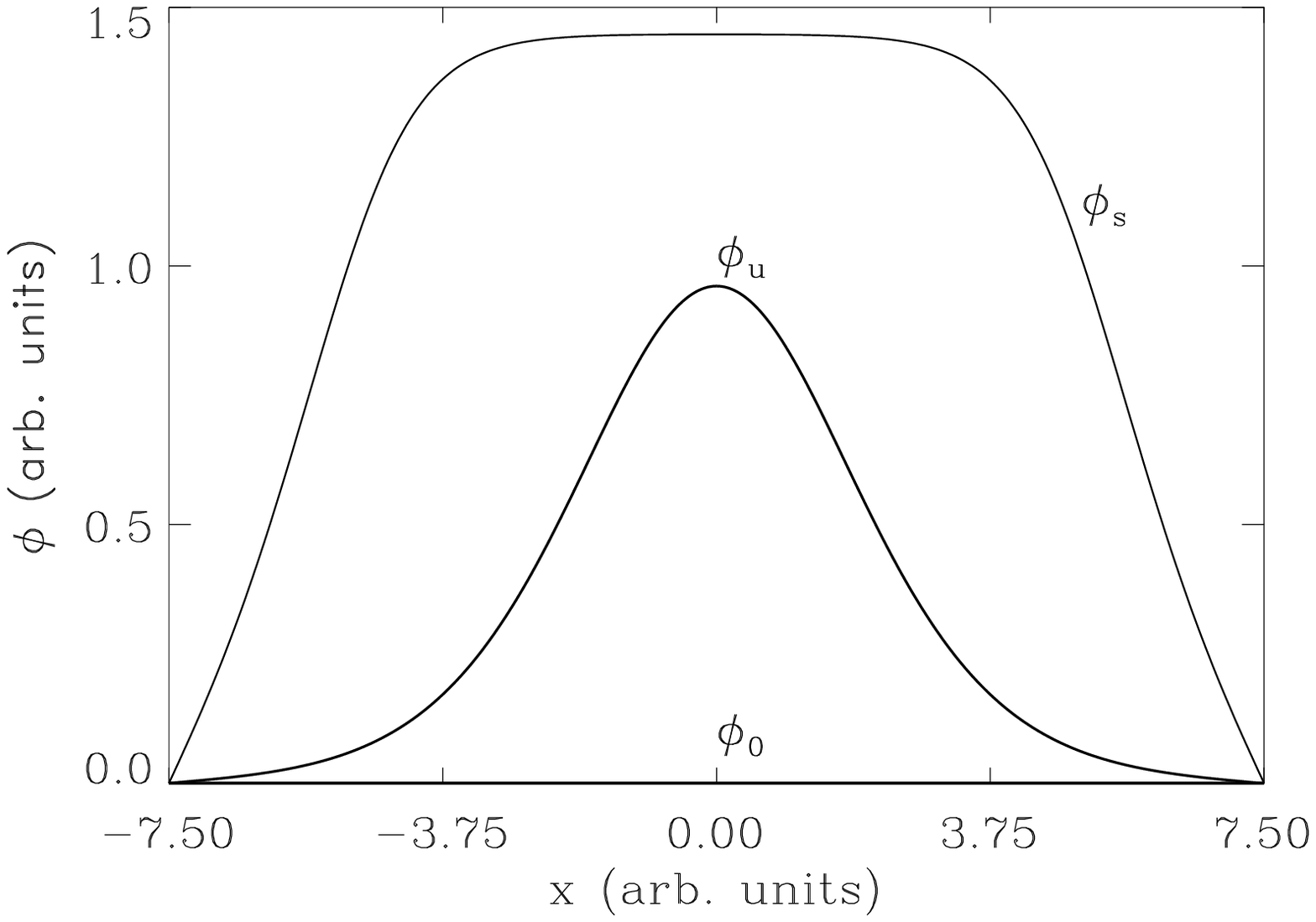}
\caption{Fixed points of the noise-sustained dynamics of the
problem. We show $\phi_0 \equiv 0$, the stable homogeneous
solution; and both nonhomogeneous patterns: the unstable (saddle)
$\phi_u$ and the stable one $\phi_s$. Here we have
$\lambda=\lambda_c \approx 0.8$, while the other parameters values
are the same as in Fig. 1}
\end{figure}

In Fig. 3 we show $V_{eff}[\phi_{st}]$ vs. $\lambda$, evaluated on
the different stationary solutions. We define $\lambda_c$ as the
value of $\lambda$ at which we have symmetrical stability, i. e.
where $V_{eff}[\phi_0]=V_{eff}[\phi_s]$.

\begin{figure}
\centering
\includegraphics[width=10cm]{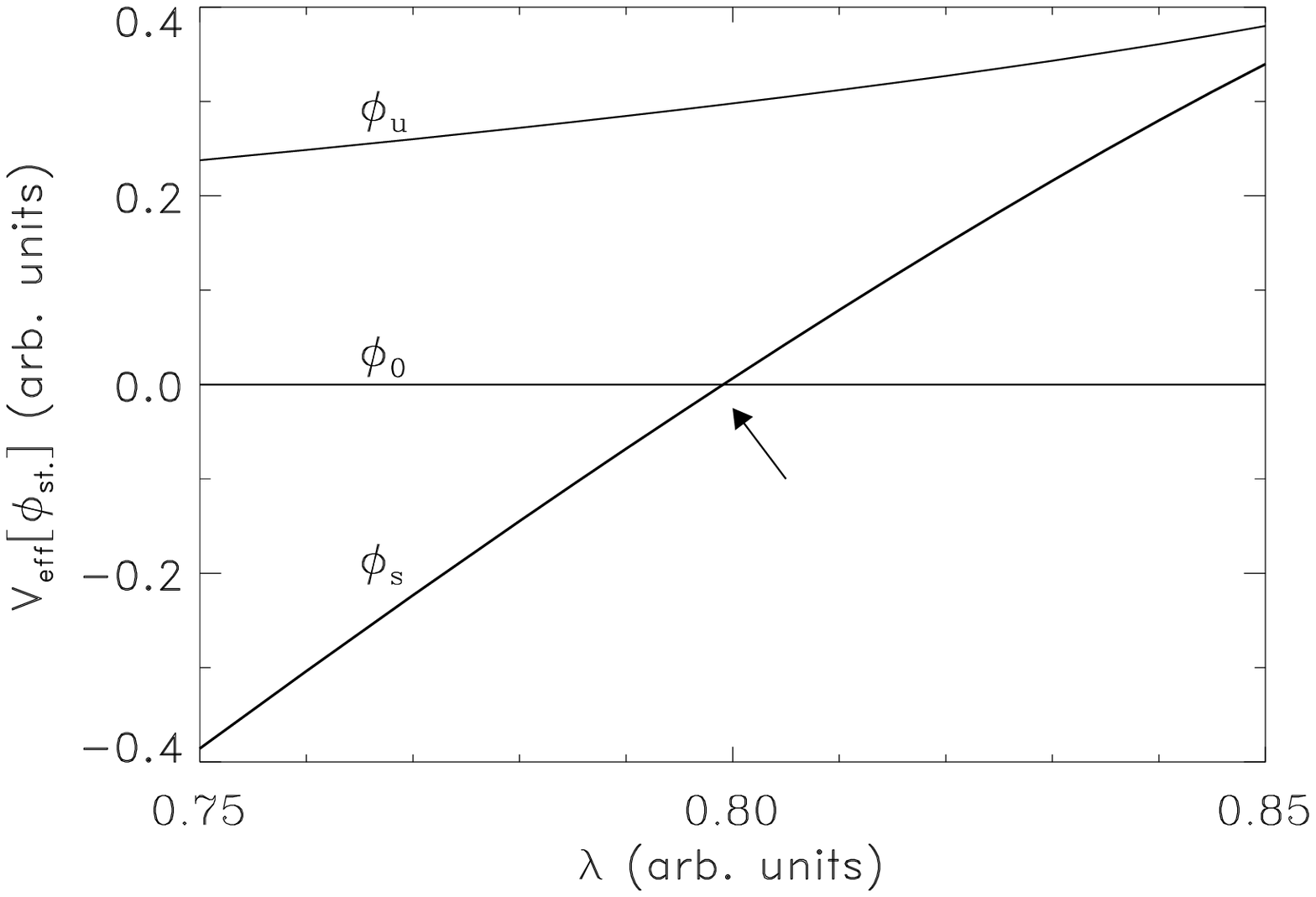}
\caption{ Nonequilibrium potential $V_{eff}[\phi_{st}]$, as a
function of $\lambda$, evaluated on the stationary patterns:
curves correspond to stable ($\phi_s$), homogeneous ($\phi_0$) and
unstable ($\phi_u$) patterns. The arrow indicate the point where
$V_{eff}[\phi_0]=V_{eff}[\phi_s]$, corresponding to $\lambda=
\lambda_c \approx 0.8$.}
\end{figure}

\section{\label{sr:sec}Stochastic resonance between structures}

We are interested in the stochastic resonance phenomena occurring
in the above described system. For a window of noise intensity the
effective dynamics of the system is bistable, corresponding to a
noise-induced nontrivial dynamics. We will resort to the so-called
\textit{two-state approximation} \cite{McNam}, all details about
the procedure and the evaluation of the SNR could be found in
\cite{extend3a}. We consider now that the system is subject, in
the adiabatic limit, to a time periodic signal of the form $b=b_0+
S(t)$  where $S(t)=\Delta b \, \sin(\omega_0 t )$. The usual way
of rocking the potential is to introduce an additive periodic
forcing (or linear periodic contribution to the potential).
However, in the present case, a small periodic variation of $b$
around $b_0$ results to be more sensitive to induce the periodic
change in the relative stability of the two attractors.

Up to first-order in the amplitude $\Delta b$ (assumed to be small
in order to have a sub-threshold periodic input) the transition
rates $W_i$ take the form
\begin{eqnarray}
\label{rates1}
W_1(t) & = & \mu_1 - \alpha_1 \, \Delta b \, \sin(\omega_0 \, t), \nonumber\\
W_2(t) & = & \mu_2 + \alpha_2 \, \Delta b \, \sin(\omega_0 \, t),
\end{eqnarray}
where the constants $\mu_{1,2}$ and $\alpha_{1,2}$ are obtained
from the Kramers-like formula for the transition rate
\cite{Hanggi}
\begin{equation}
W_{\phi_i \rightarrow \phi_j}= \frac{\lambda_+}{2 \pi} \,
\left[\frac{\det \,V_{eff}[\phi_i]}{\vert \det \,V_{eff}[\phi_u]
\vert} \right]^{1/2} \, \exp[- (V_{eff}[\phi_u]-
V_{eff}[\phi_i])/\epsilon].
\end{equation}
Here $\lambda_+$ is the unstable eigenvalue of the deterministic
flux at the relevant saddle point ($\phi_u)$ and
\begin{eqnarray}
\mu_{1,2} & = & W_{1,2} \vert_{S(t)=0}     \nonumber \\
\alpha_{1,2} & = & \mp  \frac{dW_{1,2}}{d S(t)}\vert_{S(t)=0}.
\end{eqnarray}

These results allows us to calculate the autocorrelation function,
the power spectrum and finally the SNR, that we indicate by $R$.
The details of the calculation were shown in Ref. \cite{extend3a}.
For $R$, and up to the relevant (second) order in the signal
amplitude $\Delta b$, we obtain
\begin{equation}
\label{snr} R= \, \frac{\pi}{4\,  \mu_1 \, \mu_2} \frac{(\alpha_2
\, \mu_1+\alpha_1 \, \mu_2)^2}{\mu_1 + \mu_2}= \frac{\pi}{4
\epsilon} \, \frac{\mu_1 \, \mu_2}{\mu_1+\mu_2} \, \Phi,
\end{equation}
where
\begin{equation}
\Phi=\int_{-L/2}^{L/2} dx \, \int_{\phi_0}^{\phi_s(x)} \, D(\phi')
\,  \phi'^2 \, d\phi',
\end{equation}
gives a measure of the spatial coupling strength. In our case
$\phi_0 \equiv 0$ and
\begin{equation}
\Phi=D_0 \int_{-L/2}^{L/2} dx \,  \left \{
\frac{\phi_{s}(x)}{h}-\frac{\arctan[\sqrt{h}\,\phi_{s}(x)]}{h^{3/2}}\right
\}.
\end{equation}
In Fig. 4 we show the SNR as a function of the parameter $\lambda$
(which is proportional to $\epsilon$). The existence of the
typical maximum is the characteristic fingerprint of SR. For a
window of noise intensity values, the system enhances the output
to the input periodic signal. We see that the maximum SNR occurs
at the symmetric situation, that is at $\lambda=\lambda_c$.

\begin{figure}
\centering
\includegraphics[width=10cm]{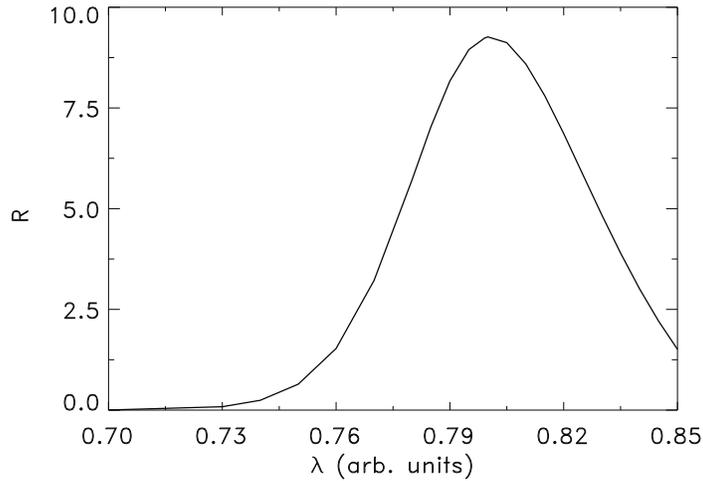}
\caption{Signal-to-noise ratio vs. $\lambda$ as results from Eq.
(\ref{snr}). Here $b_0=2$, while other parameters remain
unchanged.}
\end{figure}

A similar behavior is observed in general for a wide range of
values for $h$ and $D_0$ compatible with a bistable effective
dynamics. In particular, $\lambda_c$ is a monotonically decreasing
function of $h$, as we show in Fig. 5. For a given value of $h$, a
numerical analysis of Eq. (\ref{snr}) indicates that the maximum
of SNR take place at $\lambda_c(h)$. Note that, for a given value
of $\lambda$, $h$ appears as a additional control parameter that
allows a fine tuning of the symmetrical condition. Finally, in
Fig. 6 we show $R_c=R(\lambda_c)$ vs. $h$ in the range of values
where Kramer's formulae applies \cite{nota}.

\begin{figure}
\centering
\includegraphics[width=10cm]{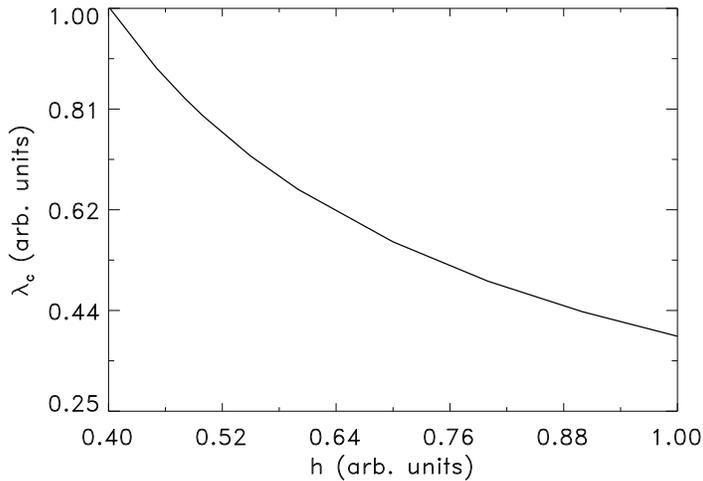}
\caption{$\lambda_c$ vs. $h$ parameter of diffusivity. For small
$h$ values $\lambda_c$, and hence the noise intensity, increase
monotonically.}
\end{figure}

\begin{figure}
\centering
\includegraphics[width=10cm]{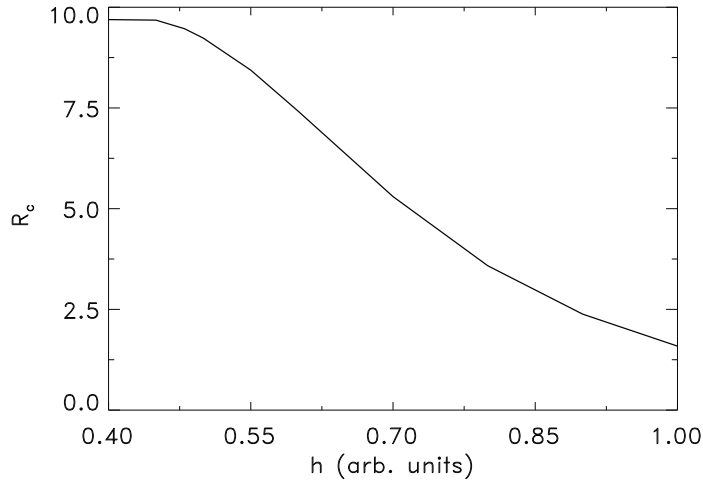}
\caption{Signal-to-noise ratio at $\lambda_c$ vs. $h$. We can see
a saturation phenomena as $h$ decreases.}
\end{figure}

\section{\label{conc}Conclusions}

The study of SR in extended or coupled systems, motivated by both,
some experimental results and the technological interest, has
recently attracted considerable attention
\cite{extend1,otros,extend2,extend2b,extend3a,extend3b}. In some
previous papers \cite{extend2,extend2b,extend3a,extend3b} we have
studied the SR phenomenon for the transition between two different
patterns, exploiting the concept of \textit{nonequilibrium
potential} \cite{GR,I0}. In this work we have analyzed the SR
phenomenon in an extended system from a different point of view,
that is studying SR between two attractors of the
\textit{noise-sustained dynamics} \cite{ga93,nsp}.

Some closely related work correspond to the so called
\textit{doubly stochastic resonance} \cite{ZKSG1}, as well as to a
previous work \cite{MGOC} that is tightly related to noise-induced
phase transitions \cite{nipt1,nipt2}. In both cases the authors
have mainly resorted to a standard mean-field approach, or to some
estimate of the effective potential. Here we adopt a different
approach, obtaining numerically the exact form of the patterns
(both the stable and unstable ones) as well as the analytical
expression of the nonequilibrium potential. In this way we were
able to obtain the transition rates and clearly quantify the SR
phenomenon by means of the SNR.

We have seen that the a nonhomogeneous spatial coupling, through
density-dependent diffusivity, changes the effective dynamics of
the system and, in agreement with \cite{extend3c}, that such
nonhomogeneous behavior could contribute to enhance the SR
phenomenon. The form of the patterns, position of the attractors,
barrier's high, explicitly depend on the noise intensity. We have
found that there are ranges or windows of noise intensities where
the phenomenon could arise (reentrance).

By considering the adiabatic limit and exploiting the two-state
approximation we have theoretically predicted the occurrence of SR
between those patterns. It is worth here remarking that it is the
same noise source the one that sustains the bistable dynamics and
induces SR for transitions among the corresponding structures. The
maximum of the SR response occurs in the symmetric case, in
agreement with the results found in \cite{extend3a,extend3b}. The
SR phenomenon is robust respect to variations of the $h$ parameter
of diffusivity, and when $h$ decreases the SNR maximum increases
and shifts toward higher $\lambda$ values. The last fact follows
from the associated shift of the noise-induced transition to
larger noise intensities which take place in the spatially
uncoupled associated system (i.e. the 0-d system resulting from
suppressing the gradient term in Eq.(\ref{pot})).

The consideration of more general forms of couplings in many
component systems will allow us to analyze SR between
noise-induced patterns in activator-inhibitor-like systems. We
will also study, within the present framework, the competence
between local and non-local spatial couplings
\cite{extend2b,extend3b}, etc. These aspects, together with Monte
Carlo simulations of the different cases, will be the subject of
further work.

\begin{acknowledgments}
The authors thanks Prof. R. Toral for fruitful discussions. H.S.
Wio acknowledges partial support from ANPCyT, Argentine, and
thanks the MECyD, Spain, for an award within the
\textit{Sabbatical Program for Visiting Professors}, and to the
Universitat de les Illes Balears for the kind hospitality extended
to him.
\end{acknowledgments}


\begin{thebibliography}{99}
\bibitem{nota} For small $h$, $\lambda_c$ increases (and hence
$\epsilon$) and Kramers formulae is (in principle) not longer
valid in this limit case.
\end{thebibliography}
\end{document}